\DeclareMathAlphabet{\mathscr}{OT1}{pzc}{m}{it}

\documentclass[prb,footinbib,twocolumn,notitlepage]{revtex4-1}
\usepackage{graphicx}
\usepackage{epsfig}
\usepackage{dcolumn}
\usepackage{bm}
\usepackage{amsmath}
\usepackage{amsbsy}
\usepackage{amssymb}
\usepackage[usenames]{color}
\usepackage{mathrsfs}
\usepackage{verbatim}
\usepackage{hyperref}
\usepackage{graphicx}

\begin{document}

\title{The paraxial approximation fails to describe the interaction of atoms with general vortex light fields}

\author{Guillermo F. Quinteiro$^1$}
\email{gfquinteiro@exa.unne.edu.ar}
\author{Christian T. Schmiegelow$^2$, Ferdinand Schmidt-Kaler$^3$}

\affiliation{
  $^1$IMIT and Departamento de F\'isica, Universidad Nacional del Nordeste, Campus Deodoro Roca, Corrientes, Argentina \\
  $^2$Departamento de F\'isica and IFIBA, FCEN, Universidad de Buenos Aires, Ciudad Universitaria, Pabell\'on I, 1428 Ciudad de Buenos Aires, Argentina \\
  $^3$QUANTUM Institut, Universitaet Mainz, Mainz, Germany}

\date{\today}

\begin{abstract} % abstract
Neglecting the electric field component along the light's propagation direction is a common practice, known as paraxial approximation. However, experimental evidence and theory on the head-on excitation of atoms by Laguerre-Gaussian beams reveal that the full vector character of the light field has to be taken into account. Optical vortices are a large family of fields, being Laguerre-Gaussian only one particular case. Now we extend the study of the applicability of the paraxial approximation to a broader set of optical vortices by considering the interaction of atoms with generalized Laguerre-Gaussian and generalized Bessel beams. We demonstrate that all vortex beams here considered look the same close to the beam axis, where the atom is placed. From this, we conclude that the paraxial approximation fails for the large set of vortex beams, and that a full understanding of the interaction of optical vortices with atoms requires the inclusion of all vector components of the electric field.
\end{abstract}

\maketitle

%%%%%%%%%%%%%%%%
%% 
\section*{Introduction}

The  paraxial approximation (PA) is the assumption that rays comprising a light beam are almost parallel to the optical axis. Mathematically, we characterize the beam parallelism by the paraxial ratio $q_r/q_z$ of radial to longitudinal wave number with $k^2=q_z^2+q_r^2$  and $q_r \propto w_0^{-1}$, or equivalently by the ratio $\lambda/w_0$ of wavelength ($\lambda$) to waist ($w_0$). The paraxial approximation can be derived from the wave equation \cite{siegman1986lasers} or Maxwell’s equations \cite{lax1975maxwell}. In the latter case, one expands the electromagnetic fields in terms of powers of $q_r/q_z$; each term in the series accounts for an increasing degree of non-paraxiality. To order zero in $q_r/q_z$, the beam is completely transverse, a common approximation in theory and experiments dealing with Transverse Electromagnetic Modes. A less stringent demand is to retain the first order in $q_r/q_z$, that leads to a non-zero component of the electric field in the direction of light's propagation.

Optical vortices (OV) or twisted light are light beams with phase singularities, carrying spin and orbital angular momentum (OAM). Research in OV spans nowadays a diversity of physics areas \cite{andrews2011str, Arikawa:17, sakamoto2016flexible, peshkov2018rayleigh, surzhykov2015interaction, oncan2018effect, bhowmik2016interaction, quinteiro2009the, shigematsu2013orb, noyan2015time, shintani2016spin, sanvitto2010persistent} with possible applications \cite{omatsu2010metal, wang2008creation, meier2007material, hernandez2017generation, pabon2017design, spinello2016radio, zhang2016self, abulikemu2016octave, miao2016orbital, seghilani2016vortex, schulze2017accelerating, woerdemann2013advanced}. In addition to the phase singularity and OAM, OV have other surprising features defying our common sense built up on plane-wave models and Gaussian beams. In particular, OV having opposite (antiparallel) orbital and spin AM are most unusual. 

In 2016, Schmiegelow et al \cite{schmiegelow2016transfer} experimentally probed quadrupole transitions between Zeeman split magnetic levels in a single $^{40}$Ca$^+$, using Gaussian and Laguerre-Gaussian (LG) --a particular type of OV-- beams, with parameters such that $\lambda/w_0=0.27$. An application of the PA then tells that the beam is transverse. However, the experimental data plus a theoretical model revealed \cite{quinteiro2017twisted} that the longitudinal component of antiparallel LG beams cannot be neglected. We thus inferred the failure of the PA for antiparallel LG fields. 

More recently we demonstrated \cite{quinteiro2019reexamination} that OV are a much larger family of beams than previously thought, that includes varieties of LG and Bessel (BB) radial modes. It is natural now to wonder if the failure of the PA found in Ref. \onlinecite{quinteiro2017twisted} for a specific antiparallel LG is a feature of general OV. We theoretically demonstrate that in fact general OV violate the PA.

%%%%%%%%%%%%%%%%
%% 
\section*{Experimental and theoretical studies of Laguerre-Gaussian beams interacting with atoms}
\label{Sec:model}

The experiment probing quadrupole transitions in a single $^{40}$Ca$^+$ (Fig 1) used Gaussian and LG beams, with OAM per photon $\hbar \ell= 0,\pm \hbar$, respectively, circular polarization (polarization vector $e_\sigma=(\hat{x}+i\sigma\hat{y})/\sqrt{2}$) $\sigma=\pm 1$, and wavelength $\lambda=0.729~ \mu$m. 
Under these conditions, the paraxial ratio is $\lambda/w_0=0.27$. As discussed above, the PA dictates the neglect of the longitudinal component (here along the $z$ axis) for modeling such a beam. Accordingly, the first theoretical model \cite{schmiegelow2016transfer} disregarded the electric field component $E_z$. This model was quite successful in reproducing the data for quadrupole transitions induced by Gaussian and parallel-momenta LG beams. However, data points for transitions induced by antiparallel-momenta LG beams could not be understood within the PA model. This motivated a reexamination of the model's assumptions, that led to the understanding that $E_z$ might play an unexpected relevant role in the latter cases. To test whether or not this is the case, the first model (PA) was contrasted to a Vector Model (VM), as we briefly describe below --for more details see Ref.\ \onlinecite{quinteiro2017twisted}. The field is given by $\mathbf E(\mathbf r,t) = \mathbf E^{(+)}(\mathbf r) e^{-i\omega t} + c. c.$ with
\begin{subequations}
\label{Eq:E_LG}
\begin{eqnarray}
\label{Eq:Eperp}
    \mathbf E_\perp^{(+)}(\mathbf r)
&=&
    ({\hat x} + i\sigma {\hat y})
    \frac{E_ 0}{\sqrt{\pi \mid\! \ell \! \mid !}w_0}
    \left (\frac {\sqrt{2}} {w_ 0} r \right)^{\mid\ell\mid}
\nonumber \\
&& \times
    e^{i\ell\varphi} e^{i k z}
\\
\label{Eq:Ez}
    E_z^{(+)} (\mathbf r)
&=&
    -i
    \left(\ell\sigma-{\mid\!\ell\!\mid}\right)
    (1-\delta_{\ell,0})
    \frac{E_ 0}{\sqrt{\pi \mid\! \ell \! \mid !}w_0}
\nonumber \\
&& \times
    \frac{\sqrt {2}}{w_0 k}
    \left (\frac {\sqrt{2}} {w_ 0} r \right)^{\mid\ell\mid-1}
    e^{i(\ell+\sigma) \varphi} e^{i k z}
\,,
\end{eqnarray}
\end{subequations}
with separated transverse $\mathbf E_\perp^{(+)}(\mathbf r)$ and longitudinal $E_z^{(+)}(\mathbf r)$ components. Note that the paraxial ratio appears only in $E_z$ as $1/(k w_0)$, with $k=2\pi/\lambda$.

\begin{figure}[h!]
  \centering
  \includegraphics[scale=.42]{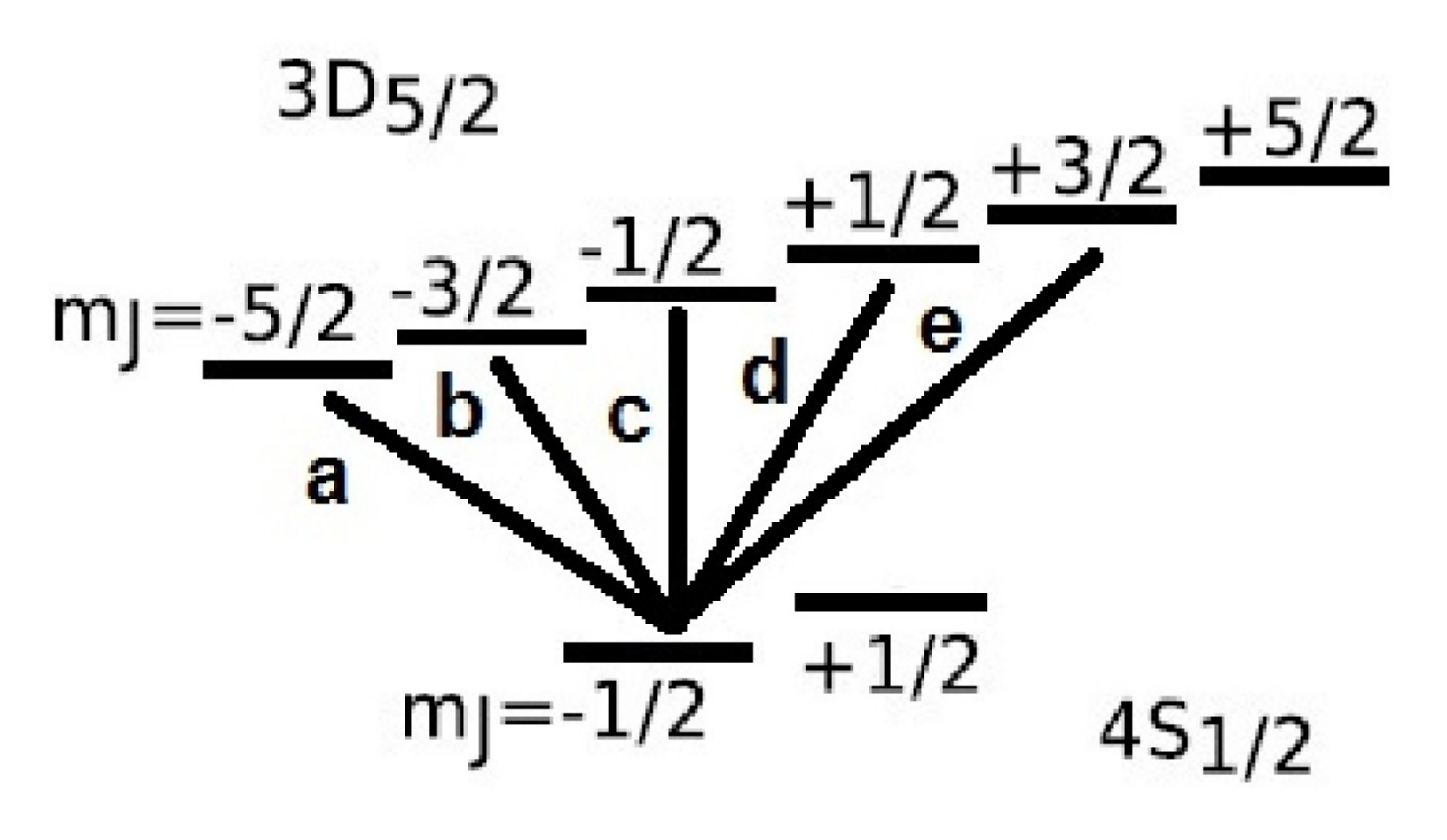}
  \caption{Zeeman-split levels in $^{40}$Ca$^+$ and transitions  induced by: (a) parallel OV, (b) Gaussian, (c) antiparallel OV, (d) Gaussian, (e) parallel OV.}
\label{fig:Trans}
\end{figure}

One can compare experiments to theory thanks to the proportionality between Rabi frequency (measured in the $^{40}$Ca$^+$ experiment) and light-matter interaction matrix element. We considered two different theoretical models: i) PA: the ion interacts solely with the transverse component of the electric field in all cases or ii) VM: the ion interacts with all three component of the electric field. Both models were compared to experimental data to draw conclusions on the validity of the PA for LG beams of the type used in Ref.\ \onlinecite{schmiegelow2016transfer}. More specifically, the theoretical procedure consisted of: i) writing down the light-matter Hamiltonian $H_I = -q \int_0^1 \mathbf r \cdot \mathbf E(u\mathbf r,t) du$; ii) separating longitudinal and transverse sections; iii) transforming from cylindrical to spherical coordinates; iv) rewriting in terms of spherical tensors; v) using Wigner-Eckard's theorem to evaluate matrix elements between initial (S-) and final (D-) states of the atom; finally, (vi) comparing ratios of matrix elements to ratios of measured Rabi frequencies. 

We tested the two different models mentioned above. Table 1 shows the results. For example, ``c/a'' is the ratio of transitions induced by an antiparallel (``c'') to a parallel (``a'') momenta LG beams, and the matrix element corresponding to “c” can be calculated either without (second row) or with (third row) $E_z$. An excellent match between theory and experiment occurs for all cases when the longitudinal component of the electric field is included (third row). In contrast, when the PA is assumed, there is agreement for ratios of transitions induced solely by parallel OV and Gaussian beams. From this results we concluded that Laguerre-Gaussian beams of the type considered violate the PA approximation.

\begin{table}[h!]
\caption{Experimental data (ratios of Rabi frequencies) and theoretical predictions (ratios of matrix elements) of the ion-OV interaction. Two different models are considered: (i) PA, that assumes $E_z=0$ even for anti-parallel OV (ii) VM, that assumes $E_z\neq 0$ for anti-parallel OV.}
\begin{tabular}{|p{2.4cm}|c|c|c|c|}
\hline & c/a & e/a & e/c & d/b \\
\cline{2-5}
\hline Experimental data & $0.95 (6)$ & $0.43 (2)$ & $0.45 (2)$ & $0.61 (3)$ \\
\hline Paraxial \mbox{Approximation} model  $\left[E_z(r)=0 \right]$ & $0.32$ & $0.45$ & $1.41$ & $0.71$ \\
\hline Vector Model $\left[E_{z}(\mathrm{r}) \neq 0\right]$ & $0.95$ & $0.45$ & $0.47$ & $0.71$ \\
\hline
\end{tabular}
\end{table}

The question arises, if this failure of PA is a general feature of vortex beams.

%%%%%%%%%%%%%%%%
%% 
\section*{Generalized Optical Vortices}
\label{Sec:LargeFamily}

Optical vortices are a family of fields, characterized by phase singularities in one or more components. Instances of OV are: Laguerre-Gauss, Bessel, Mathieu, azimuthally polarized, etc.   

Recently \cite{quinteiro2019reexamination} we identified the existence of a large group of OV containing fields with varying degrees of relative strengths of electric to magnetic fields, parametrized by a real number $\gamma$. For example, a beam with $\gamma=1$ has a strong magnetic field at the phase singularity \cite{quinteiro2017formulation}. On the other side, one finds fields with strong electric components for $\gamma=0$. In between, a more common OV \cite{bliokh2010angular, li2009spin} with features closer to non-vortex beams is found at $\gamma = [1+\omega/(c q_z)]^{-1}$.
The general expression for the OV for all $\gamma$ are derived in Ref. \cite{quinteiro2019reexamination}, here we present the electric field
\begin{eqnarray}
\label{Eq:gamaE}
\mathbf{E}^{(\gamma)}(\mathbf{r}, t)
&=& 
i\left[\omega A_{x}+\bar{\gamma} \frac{c^{2}}{\omega} \partial_{x}\left(\nabla_{\perp} \cdot \mathbf{A}_{\perp}\right)\right] \hat{x}
\nonumber\\
&& 
+i\left[\omega A_{y}+\bar{\gamma} \frac{c^{2}}{\omega} \partial_{y}\left(\nabla_{\perp} \cdot \mathbf{A}_{\perp}\right)\right] \hat{y}
\nonumber\\
&&
-\left[\gamma \frac{\omega}{q_{z}}+\bar{\gamma} \frac{c^{2} q_{z}}{\omega}\right]\left(\nabla_{\perp} \cdot \mathbf{A}_{\perp}\right) \hat{z}
\end{eqnarray}
where $\mathbf A_i=\mathbf A_i(\mathbf r, t)$ and $A_\perp$ is the transverse part of the vector potential (either LG or BB) and $\bar{\gamma}=1-\gamma$. 

In order to model head-on ion-OV interaction experiments --such as that from Ref.\ \onlinecite{schmiegelow2016transfer}-- we approximate the field Eq.\ \ref{Eq:gamaE} up to first order in the paraxial ratio $q_r/q_z$, and look for its values in the neighborhood of $r=0$ (where the atom is placed), keeping up to first order in $q_r r$. 
A lengthy but straightforward calculation reveals that all LG and Bessel beams of the type described by Eq.\ \ref{Eq:gamaE} with $|\ell|=1$ are of the same form 
\begin{eqnarray}
&&\hspace{-7mm}
\sigma=+1: \tilde{\mathbf{E}}^{(\gamma)}(\mathbf{r})
=
E_{0} \left( r e^{i \ell \varphi} \mathbf{e}_{+} 
+i \frac{\sqrt{2}}{q_{z}} \mathbf{e}_{z} \delta_{\ell,-1}\right)
\label{Eq:gamaEAprox}
\\
&&\hspace{-7mm}
\sigma=-1: \tilde{\mathbf{E}}^{(\gamma)}(\mathbf{r})
=
E_{0}\left(r e^{i \ell \varphi} \mathbf{e}_{-}
+i \frac{\sqrt{2}}{q_{z}} \mathbf{e}_{z} \delta_{\ell,+1}\right)
\label{Eq:gamaEAprox2}
\end{eqnarray}
irrespective of the value of $\gamma$. In Eq. (1) $\mathbf{E}=\tilde{\mathbf{E}} \exp \left[i\left(q_{z} z-\omega t\right)\right]$,  with $\{r,\varphi,z\}$ cylindrical coordinates. To this level of approximation, one clearly sees that parallel OV (first equation) are transverse.

We have reached the main point of this report: Equations \ref{Eq:gamaEAprox}-\ref{Eq:gamaEAprox2} has the same structure as Eqs.\ \ref{Eq:Eperp}-\ref{Eq:Ez}; that is, the electric field of all general OV here considered are indistinguishable when $q_r/q_z$ is small and one looks close to the phase singularity. Therefore, any antiparallel OV under the same conditions stated above will yield the same results as presented in Table 1. We conclude that the PA fails {\bf in general} to describe the interaction of antiparallel $|\ell|=1$ OV with atoms.

%%%%%%%%%%%%%%%%
%% 
\section*{Conclusions}

We have demonstrated that the PA fails not only for LG beams but also for a large family of OV -including LG and Bessel- that have antiparallel spin and orbital angular momenta. All these beams share the same mathematical structure close to the phase singularity and, based on a conclusive experiment plus theory, the full vector character of the light field has to be taken into account, such that the longitudial component of the electrical field is correctly included and the OV-atom interaction is properly described. Our findings have practical consequences in new applications that make use of light-matter interaction. Examples from AMO physics include the excitation of single trapped ions, e.g. for quantum optical experiments of for accurate frequency standards or of single trapped atoms. In the field of solid state application we see the excitions of magnetic transitions in impurities such as Eu$^{+3}$ or quantum dots as highly relevant cases.

\section*{Acknowledgements}
  GFQ thanks the ONRG for financial support through NICOP  N62909-18-1-2090.

% ---- bibliography  


\begin{thebibliography}{32}
\ifx \bisbn   \undefined \def \bisbn  #1{ISBN #1}\fi
\ifx \binits  \undefined \def \binits#1{#1}\fi
\ifx \bauthor  \undefined \def \bauthor#1{#1}\fi
\ifx \batitle  \undefined \def \batitle#1{#1}\fi
\ifx \bjtitle  \undefined \def \bjtitle#1{#1}\fi
\ifx \bvolume  \undefined \def \bvolume#1{\textbf{#1}}\fi
\ifx \byear  \undefined \def \byear#1{#1}\fi
\ifx \bissue  \undefined \def \bissue#1{#1}\fi
\ifx \bfpage  \undefined \def \bfpage#1{#1}\fi
\ifx \blpage  \undefined \def \blpage #1{#1}\fi
\ifx \burl  \undefined \def \burl#1{\textsf{#1}}\fi
\ifx \doiurl  \undefined \def \doiurl#1{\textsf{#1}}\fi
\ifx \betal  \undefined \def \betal{\textit{et al.}}\fi
\ifx \binstitute  \undefined \def \binstitute#1{#1}\fi
\ifx \binstitutionaled  \undefined \def \binstitutionaled#1{#1}\fi
\ifx \bctitle  \undefined \def \bctitle#1{#1}\fi
\ifx \beditor  \undefined \def \beditor#1{#1}\fi
\ifx \bpublisher  \undefined \def \bpublisher#1{#1}\fi
\ifx \bbtitle  \undefined \def \bbtitle#1{#1}\fi
\ifx \bedition  \undefined \def \bedition#1{#1}\fi
\ifx \bseriesno  \undefined \def \bseriesno#1{#1}\fi
\ifx \blocation  \undefined \def \blocation#1{#1}\fi
\ifx \bsertitle  \undefined \def \bsertitle#1{#1}\fi
\ifx \bsnm \undefined \def \bsnm#1{#1}\fi
\ifx \bsuffix \undefined \def \bsuffix#1{#1}\fi
\ifx \bparticle \undefined \def \bparticle#1{#1}\fi
\ifx \barticle \undefined \def \barticle#1{#1}\fi
\ifx \bconfdate \undefined \def \bconfdate #1{#1}\fi
\ifx \botherref \undefined \def \botherref #1{#1}\fi
\ifx \url \undefined \def \url#1{\textsf{#1}}\fi
\ifx \bchapter \undefined \def \bchapter#1{#1}\fi
\ifx \bbook \undefined \def \bbook#1{#1}\fi
\ifx \bcomment \undefined \def \bcomment#1{#1}\fi
\ifx \oauthor \undefined \def \oauthor#1{#1}\fi
\ifx \citeauthoryear \undefined \def \citeauthoryear#1{#1}\fi
\ifx \endbibitem  \undefined \def \endbibitem {}\fi
\ifx \bconflocation  \undefined \def \bconflocation#1{#1}\fi
\ifx \arxivurl  \undefined \def \arxivurl#1{\textsf{#1}}\fi
\csname PreBibitemsHook\endcsname

%%% 1
\bibitem{siegman1986lasers}
\begin{botherref}
\oauthor{\bsnm{Siegman}, \binits{A.E.}}:
Lasers. mill valley, ca: Univ.
Science,
298--301
(1986)
\end{botherref}
\endbibitem

%%% 2
\bibitem{lax1975maxwell}
\begin{barticle}
\bauthor{\bsnm{Lax}, \binits{M.}},
\bauthor{\bsnm{Louisell}, \binits{W.H.}},
\bauthor{\bsnm{McKnight}, \binits{W.B.}}:
\batitle{From maxwell to paraxial wave optics}.
\bjtitle{Physical Review A}
\bvolume{11}(\bissue{4}),
\bfpage{1365}
(\byear{1975})
\end{barticle}
\endbibitem

%%% 3
\bibitem{andrews2011str}
\begin{bbook}
\bauthor{\bsnm{Andrews}, \binits{D.L.}}:
\bbtitle{Structured Light and Its Applications: An Introduction to
  Phase-structured Beams and Nanoscale Optical Forces}.
\bpublisher{Academic Press}
(\byear{2008})
\end{bbook}
\endbibitem

%%% 4
\bibitem{Arikawa:17}
\begin{barticle}
\bauthor{\bsnm{Arikawa}, \binits{T.}},
\bauthor{\bsnm{Morimoto}, \binits{S.}},
\bauthor{\bsnm{Tanaka}, \binits{K.}}:
\batitle{Focusing light with orbital angular momentum by circular array
  antenna}.
\bjtitle{Optics Express}
\bvolume{25}(\bissue{12}),
\bfpage{13728}--\blpage{13735}
(\byear{2017}).
\end{barticle}
\endbibitem

%%% 5
\bibitem{sakamoto2016flexible}
\begin{barticle}
\bauthor{\bsnm{Sakamoto}, \binits{M.}},
\bauthor{\bsnm{Sasaki}, \binits{T.}},
\bauthor{\bsnm{Tien}, \binits{T.M.}},
\bauthor{\bsnm{Noda}, \binits{K.}},
\bauthor{\bsnm{Kawatsuki}, \binits{N.}},
\bauthor{\bsnm{Ono}, \binits{H.}}:
\batitle{Flexible and achromatic generation of optical vortices by use of
  vector beam recorded functionalized liquid crystals}.
\bjtitle{Applied Optics}
\bvolume{55}(\bissue{36}),
\bfpage{10427}--\blpage{10434}
(\byear{2016})
\end{barticle}
\endbibitem

%%% 6
\bibitem{peshkov2018rayleigh}
\begin{barticle}
\bauthor{\bsnm{Peshkov}, \binits{A.}},
\bauthor{\bsnm{Volotka}, \binits{A.}},
\bauthor{\bsnm{Surzhykov}, \binits{A.}},
\bauthor{\bsnm{Fritzsche}, \binits{S.}}:
\batitle{Rayleigh scattering of twisted light by hydrogenlike ions}.
\bjtitle{Physical Review A}
\bvolume{97}(\bissue{2}),
\bfpage{023802}
(\byear{2018})
\end{barticle}
\endbibitem

%%% 7
\bibitem{surzhykov2015interaction}
\begin{barticle}
\bauthor{\bsnm{Surzhykov}, \binits{A.}},
\bauthor{\bsnm{Seipt}, \binits{D.}},
\bauthor{\bsnm{Serbo}, \binits{V.}},
\bauthor{\bsnm{Fritzsche}, \binits{S.}}:
\batitle{Interaction of twisted light with many-electron atoms and ions}.
\bjtitle{Physical Review A}
\bvolume{91}(\bissue{1}),
\bfpage{013403}
(\byear{2015})
\end{barticle}
\endbibitem

%%% 8
\bibitem{oncan2018effect}
\begin{botherref}
\oauthor{\bsnm{{\"O}ncan}, \binits{M.}},
\oauthor{\bsnm{Ko{\c{c}}}, \binits{F.}},
\oauthor{\bsnm{Dereli}, \binits{D.B.}},
\oauthor{\bsnm{K{\"o}ksal}, \binits{K.}}:
The effect of radial and angular profiles of twisted laser beam on coronene
  molecule located off the optical axis.
Computational and Theoretical Chemistry
(2018)
\end{botherref}
\endbibitem

%%% 9
\bibitem{bhowmik2016interaction}
\begin{barticle}
\bauthor{\bsnm{Bhowmik}, \binits{A.}},
\bauthor{\bsnm{Mondal}, \binits{P.K.}},
\bauthor{\bsnm{Majumder}, \binits{S.}},
\bauthor{\bsnm{Deb}, \binits{B.}}:
\batitle{Interaction of atom with nonparaxial laguerre-gaussian beam: Forming
  superposition of vortex states in bose-einstein condensates}.
\bjtitle{Physical Review A}
\bvolume{93}(\bissue{6}),
\bfpage{063852}
(\byear{2016})
\end{barticle}
\endbibitem

%%% 10
\bibitem{quinteiro2009the}
\begin{barticle}
\bauthor{\bsnm{Quinteiro}, \binits{G.F.}},
\bauthor{\bsnm{Tamborenea}, \binits{P.I.}}:
\batitle{Theory of the optical absorption of light carrying orbital angular
  momentum by semiconductors}.
\bjtitle{Europhysics Letters}
\bvolume{85},
\bfpage{47001}
(\byear{2009})
\end{barticle}
\endbibitem

%%% 11
\bibitem{shigematsu2013orb}
\begin{barticle}
\bauthor{\bsnm{Shigematsu}, \binits{K.}},
\bauthor{\bsnm{Toda}, \binits{Y.}},
\bauthor{\bsnm{Yamane}, \binits{K.}},
\bauthor{\bsnm{Morita}, \binits{R.}}:
\batitle{Orbital angular momentum spectral dynamics of gan excitons excited by
  optical vortices}.
\bjtitle{Japanese Journal of Applied Physics}
\bvolume{52},
\bfpage{08}--\blpage{08}
(\byear{2013})
\end{barticle}
\endbibitem

%%% 12
\bibitem{noyan2015time}
\begin{barticle}
\bauthor{\bsnm{Noyan}, \binits{M.A.}},
\bauthor{\bsnm{Kikkawa}, \binits{J.M.}}:
\batitle{Time-resolved orbital angular momentum spectroscopy}.
\bjtitle{Applied Physics Letters}
\bvolume{107}(\bissue{3}),
\bfpage{032406}
(\byear{2015})
\end{barticle}
\endbibitem

%%% 13
\bibitem{shintani2016spin}
\begin{barticle}
\bauthor{\bsnm{Shintani}, \binits{K.}},
\bauthor{\bsnm{Taguchi}, \binits{K.}},
\bauthor{\bsnm{Tanaka}, \binits{Y.}},
\bauthor{\bsnm{Kawaguchi}, \binits{Y.}}:
\batitle{Spin and charge transport induced by a twisted light beam on the
  surface of a topological insulator}.
\bjtitle{Physical Review B}
\bvolume{93}(\bissue{19}),
\bfpage{195415}
(\byear{2016})
\end{barticle}
\endbibitem

%%% 14
\bibitem{sanvitto2010persistent}
\begin{barticle}
\bauthor{\bsnm{Sanvitto}, \binits{D.}},
\bauthor{\bsnm{Marchetti}, \binits{F.}},
\bauthor{\bsnm{Szyma{\'n}ska}, \binits{M.}},
\bauthor{\bsnm{Tosi}, \binits{G.}},
\bauthor{\bsnm{Baudisch}, \binits{M.}},
\bauthor{\bsnm{Laussy}, \binits{F.}},
\bauthor{\bsnm{Krizhanovskii}, \binits{D.}},
\bauthor{\bsnm{Skolnick}, \binits{M.}},
\bauthor{\bsnm{Marrucci}, \binits{L.}},
\bauthor{\bsnm{Lemaitre}, \binits{A.}}, \betal:
\batitle{Persistent currents and quantized vortices in a polariton superfluid}.
\bjtitle{Nature Physics}
\bvolume{6}(\bissue{7}),
\bfpage{527}--\blpage{533}
(\byear{2010})
\end{barticle}
\endbibitem

%%% 15
\bibitem{omatsu2010metal}
\begin{barticle}
\bauthor{\bsnm{Omatsu}, \binits{T.}},
\bauthor{\bsnm{Chujo}, \binits{K.}},
\bauthor{\bsnm{Miyamoto}, \binits{K.}},
\bauthor{\bsnm{Okida}, \binits{M.}},
\bauthor{\bsnm{Nakamura}, \binits{K.}},
\bauthor{\bsnm{Aoki}, \binits{N.}},
\bauthor{\bsnm{Morita}, \binits{R.}}:
\batitle{Metal microneedle fabrication using twisted light with spin}.
\bjtitle{Optics Express}
\bvolume{18}(\bissue{17}),
\bfpage{17967}--\blpage{17973}
(\byear{2010})
\end{barticle}
\endbibitem

%%% 16
\bibitem{wang2008creation}
\begin{barticle}
\bauthor{\bsnm{Wang}, \binits{H.}},
\bauthor{\bsnm{Shi}, \binits{L.}},
\bauthor{\bsnm{Lukyanchuk}, \binits{B.}},
\bauthor{\bsnm{Sheppard}, \binits{C.}},
\bauthor{\bsnm{Chong}, \binits{C.T.}}:
\batitle{Creation of a needle of longitudinally polarized light in vacuum using
  binary optics}.
\bjtitle{Nature Photonics}
\bvolume{2}(\bissue{8}),
\bfpage{501}--\blpage{505}
(\byear{2008})
\end{barticle}
\endbibitem

%%% 17
\bibitem{meier2007material}
\begin{barticle}
\bauthor{\bsnm{Meier}, \binits{M.}},
\bauthor{\bsnm{Romano}, \binits{V.}},
\bauthor{\bsnm{Feurer}, \binits{T.}}:
\batitle{Material processing with pulsed radially and azimuthally polarized
  laser radiation}.
\bjtitle{Applied Physics A: Materials Science \& Processing}
\bvolume{86}(\bissue{3}),
\bfpage{329}--\blpage{334}
(\byear{2007})
\end{barticle}
\endbibitem

%%% 18
\bibitem{hernandez2017generation}
\begin{bchapter}
\bauthor{\bsnm{Hern{\'a}ndez-Garc{\'\i}a}, \binits{C.}},
\bauthor{\bsnm{Vieira}, \binits{J.}},
\bauthor{\bsnm{Mendon{\c{c}}a}, \binits{J.T.}},
\bauthor{\bsnm{Rego}, \binits{L.}},
\bauthor{\bsnm{San~Rom{\'a}n}, \binits{J.}},
\bauthor{\bsnm{Plaja}, \binits{L.}},
\bauthor{\bsnm{Ribic}, \binits{P.R.}},
\bauthor{\bsnm{Gauthier}, \binits{D.}},
\bauthor{\bsnm{Pic{\'o}n}, \binits{A.}}:
\bctitle{Generation and applications of extreme-ultraviolet vortices}.
In: \bbtitle{Photonics},
vol. \bseriesno{4},
p. \bfpage{28}
(\byear{2017}).
\bcomment{Multidisciplinary Digital Publishing Institute}
\end{bchapter}
\endbibitem

%%% 19
\bibitem{pabon2017design}
\begin{barticle}
\bauthor{\bsnm{Pabon}, \binits{D.}},
\bauthor{\bsnm{Ledesma}, \binits{S.}},
\bauthor{\bsnm{Quinteiro}, \binits{G.}},
\bauthor{\bsnm{Capeluto}, \binits{M.}}:
\batitle{Design of a compact device to generate and test beams with orbital
  angular momentum in the euv}.
\bjtitle{Applied Optics}
\bvolume{56}(\bissue{29}),
\bfpage{8048}--\blpage{8054}
(\byear{2017})
\end{barticle}
\endbibitem

%%% 20
\bibitem{spinello2016radio}
\begin{barticle}
\bauthor{\bsnm{Spinello}, \binits{F.}},
\bauthor{\bsnm{Someda}, \binits{C.G.}},
\bauthor{\bsnm{Ravanelli}, \binits{R.A.}},
\bauthor{\bsnm{Mari}, \binits{E.}},
\bauthor{\bsnm{Parisi}, \binits{G.}},
\bauthor{\bsnm{Tamburini}, \binits{F.}},
\bauthor{\bsnm{Romanato}, \binits{F.}},
\bauthor{\bsnm{Coassini}, \binits{P.}},
\bauthor{\bsnm{Oldoni}, \binits{M.}}:
\batitle{Radio channel multiplexing with superpositions of opposite-sign oam
  modes}.
\bjtitle{AEU-International Journal of Electronics and Communications}
\bvolume{70}(\bissue{8}),
\bfpage{990}--\blpage{997}
(\byear{2016})
\end{barticle}
\endbibitem

%%% 21
\bibitem{zhang2016self}
\begin{barticle}
\bauthor{\bsnm{Zhang}, \binits{Y.}},
\bauthor{\bsnm{Yu}, \binits{H.}},
\bauthor{\bsnm{Zhang}, \binits{H.}},
\bauthor{\bsnm{Xu}, \binits{X.}},
\bauthor{\bsnm{Xu}, \binits{J.}},
\bauthor{\bsnm{Wang}, \binits{J.}}:
\batitle{Self-mode-locked laguerre-gaussian beam with staged topological charge
  by thermal-optical field coupling}.
\bjtitle{Optics Express}
\bvolume{24}(\bissue{5}),
\bfpage{5514}--\blpage{5522}
(\byear{2016})
\end{barticle}
\endbibitem

%%% 22
\bibitem{abulikemu2016octave}
\begin{barticle}
\bauthor{\bsnm{Abulikemu}, \binits{A.}},
\bauthor{\bsnm{Yusufu}, \binits{T.}},
\bauthor{\bsnm{Mamuti}, \binits{R.}},
\bauthor{\bsnm{Araki}, \binits{S.}},
\bauthor{\bsnm{Miyamoto}, \binits{K.}},
\bauthor{\bsnm{Omatsu}, \binits{T.}}:
\batitle{Octave-band tunable optical vortex parametric oscillator}.
\bjtitle{Optics Express}
\bvolume{24}(\bissue{14}),
\bfpage{15204}--\blpage{15211}
(\byear{2016})
\end{barticle}
\endbibitem

%%% 23
\bibitem{miao2016orbital}
\begin{barticle}
\bauthor{\bsnm{Miao}, \binits{P.}},
\bauthor{\bsnm{Zhang}, \binits{Z.}},
\bauthor{\bsnm{Sun}, \binits{J.}},
\bauthor{\bsnm{Walasik}, \binits{W.}},
\bauthor{\bsnm{Longhi}, \binits{S.}},
\bauthor{\bsnm{Litchinitser}, \binits{N.M.}},
\bauthor{\bsnm{Feng}, \binits{L.}}:
\batitle{Orbital angular momentum microlaser}.
\bjtitle{Science}
\bvolume{353}(\bissue{6298}),
\bfpage{464}--\blpage{467}
(\byear{2016})
\end{barticle}
\endbibitem

%%% 24
\bibitem{seghilani2016vortex}
\begin{barticle}
\bauthor{\bsnm{Seghilani}, \binits{M.S.}},
\bauthor{\bsnm{Myara}, \binits{M.}},
\bauthor{\bsnm{Sellahi}, \binits{M.}},
\bauthor{\bsnm{Legratiet}, \binits{L.}},
\bauthor{\bsnm{Sagnes}, \binits{I.}},
\bauthor{\bsnm{Beaudoin}, \binits{G.}},
\bauthor{\bsnm{Lalanne}, \binits{P.}},
\bauthor{\bsnm{Garnache}, \binits{A.}}:
\batitle{Vortex laser based on iii-v semiconductor metasurface: direct
  generation of coherent laguerre-gauss modes carrying controlled orbital
  angular momentum}.
\bjtitle{Scientific Reports}
\bvolume{6},
\bfpage{38156}
(\byear{2016})
\end{barticle}
\endbibitem

%%% 25
\bibitem{schulze2017accelerating}
\begin{botherref}
\oauthor{\bsnm{Schulze}, \binits{D.}},
\oauthor{\bsnm{Thakur}, \binits{A.}},
\oauthor{\bsnm{Moskalenko}, \binits{A.S.}},
\oauthor{\bsnm{Berakdar}, \binits{J.}}:
Accelerating, guiding, and sub-wavelength trapping of neutral atoms with
  tailored optical vortices.
Annalen der Physik
\textbf{529}(5)
(2017)
\end{botherref}
\endbibitem

%%% 26
\bibitem{woerdemann2013advanced}
\begin{barticle}
\bauthor{\bsnm{Woerdemann}, \binits{M.}},
\bauthor{\bsnm{Alpmann}, \binits{C.}},
\bauthor{\bsnm{Esseling}, \binits{M.}},
\bauthor{\bsnm{Denz}, \binits{C.}}:
\batitle{Advanced optical trapping by complex beam shaping}.
\bjtitle{Laser \& Photonics Reviews}
\bvolume{7}(\bissue{6}),
\bfpage{839}--\blpage{854}
(\byear{2013})
\end{barticle}
\endbibitem

%%% 27
\bibitem{schmiegelow2016transfer}
\begin{barticle}
\bauthor{\bsnm{Schmiegelow}, \binits{C.T.}},
\bauthor{\bsnm{Schulz}, \binits{J.}},
\bauthor{\bsnm{Kaufmann}, \binits{H.}},
\bauthor{\bsnm{Ruster}, \binits{T.}},
\bauthor{\bsnm{Poschinger}, \binits{U.G.}},
\bauthor{\bsnm{Schmidt-Kaler}, \binits{F.}}:
\batitle{Transfer of optical orbital angular momentum to a bound electron}.
\bjtitle{Nature Communications}
\bvolume{7},
\bfpage{12998}
(\byear{2016})
\end{barticle}
\endbibitem

%%% 28
\bibitem{quinteiro2017twisted}
\begin{barticle}
\bauthor{\bsnm{Quinteiro}, \binits{G.F.}},
\bauthor{\bsnm{Schmidt-Kaler}, \binits{F.}},
\bauthor{\bsnm{Schmiegelow}, \binits{C.T.}}:
\batitle{Twisted-light--ion interaction: The role of longitudinal fields}.
\bjtitle{Physical Review Letters}
\bvolume{119}(\bissue{25}),
\bfpage{253203}
(\byear{2017})
\end{barticle}
\endbibitem

%%% 29
\bibitem{quinteiro2019reexamination}
\begin{barticle}
\bauthor{\bsnm{Quinteiro}, \binits{G.}},
\bauthor{\bsnm{Schmiegelow}, \binits{C.}},
\bauthor{\bsnm{Reiter}, \binits{D.}},
\bauthor{\bsnm{Kuhn}, \binits{T.}}:
\batitle{Reexamination of bessel beams: A generalized scheme to derive optical
  vortices}.
\bjtitle{Physical Review A}
\bvolume{99}(\bissue{2}),
\bfpage{023845}
(\byear{2019})
\end{barticle}
\endbibitem

%%% 30
\bibitem{quinteiro2017formulation}
\begin{barticle}
\bauthor{\bsnm{Quinteiro}, \binits{G.F.}},
\bauthor{\bsnm{Reiter}, \binits{D.}},
\bauthor{\bsnm{Kuhn}, \binits{T.}}:
\batitle{Formulation of the twisted-light--matter interaction at the phase
  singularity: beams with strong magnetic fields}.
\bjtitle{Physical Review A}
\bvolume{95}(\bissue{1}),
\bfpage{012106}
(\byear{2017})
\end{barticle}
\endbibitem

%%% 31
\bibitem{bliokh2010angular}
\begin{barticle}
\bauthor{\bsnm{Bliokh}, \binits{K.Y.}},
\bauthor{\bsnm{Alonso}, \binits{M.A.}},
\bauthor{\bsnm{Ostrovskaya}, \binits{E.A.}},
\bauthor{\bsnm{Aiello}, \binits{A.}}:
\batitle{Angular momenta and spin-orbit interaction of nonparaxial light in
  free space}.
\bjtitle{Physical Review A}
\bvolume{82}(\bissue{6}),
\bfpage{063825}
(\byear{2010})
\end{barticle}
\endbibitem

%%% 32
\bibitem{li2009spin}
\begin{barticle}
\bauthor{\bsnm{Li}, \binits{C.-F.}}:
\batitle{Spin and orbital angular momentum of a class of nonparaxial light
  beams having a globally defined polarization}.
\bjtitle{Physical Review A}
\bvolume{80}(\bissue{6}),
\bfpage{063814}
(\byear{2009})
\end{barticle}
\endbibitem

\end{thebibliography}
\end{document}